\documentclass[twoside,11pt]{article}
\usepackage{pgm}
\usepackage[utf8]{inputenc}
\inputencoding{utf8}

\usepackage{hyperref,color,soul,booktabs}
\setulcolor{blue}
\newcommand{\BibTeX}{\textsc{B\kern-0.1emi\kern-0.017emb}\kern-0.15em\TeX}

\usepackage{colortbl}

\usepackage{amssymb}
\usepackage{multirow}
\usepackage{booktabs,subcaption,amsfonts,dcolumn}
\newcolumntype{d}[1]{D..{#1}}
\newcommand{\specialcell}[2][c]{\begin{tabular}[#1]{@{}c@{}}#2\end{tabular}}
\usepackage{algorithmic}
\usepackage{subcaption}
\usepackage{graphicx}
\usepackage{amsmath}
\usepackage{amssymb}
\usepackage{url}
\usepackage{geometry}
\usepackage{color}
\usepackage{booktabs}
\usepackage[ruled,vlined]{algorithm2e}
\usepackage{hyperref}

\definecolor{RED}{rgb}{1,0,0}
\definecolor{ORANGE}{rgb}{1,0.5,0}
\definecolor{BLUE}{rgb}{0,0,1}
\newcommand{\FIXME}[1]{{\color{RED}{\textbf{FIX}: #1}}}

\usepackage{pgf, tikz} 
\usetikzlibrary{arrows,automata,fit}
\usetikzlibrary{shapes}

\usepackage{makecell}
\usepackage{bm}
\usepackage{lscape}

\usepackage{verbatim}

\usepackage{algorithmic}
\ifpdf
\DeclareGraphicsExtensions{.eps,.pdf,.png,.jpg}
\else
\DeclareGraphicsExtensions{.eps}
\fi

\ShortHeadings{The YODO algorithm}{Ballester-Ripoll and Leonelli}

\begin{document}

\title{The YODO algorithm: An efficient computational framework for sensitivity analysis in Bayesian networks}
\author{\Name{Rafael Ballester-Ripoll} \Email{rafael.ballester@ie.edu}\and
   \Name{Manuele Leonelli} \Email{manuele.leonelli@ie.edu}\\
   \addr School of Science and Technology, IE University, Madrid, Spain}

\maketitle

\begin{abstract}
Sensitivity analysis measures the influence of a Bayesian network’s parameters on a quantity of interest defined by the network, such as the probability of a variable taking a specific value. Various sensitivity measures have been defined to quantify such influence, most commonly some function of the quantity of interest’s partial derivative with respect to the network’s conditional probabilities. However, computing these measures in large networks with thousands of parameters can become computationally very expensive. We propose an algorithm combining automatic differentiation and exact inference to efficiently calculate the sensitivity measures in a single pass. It first marginalizes the whole network once, using e.g. variable elimination, and then backpropagates this operation to obtain the gradient with respect to all input parameters. Our method can be used for one-way and multi-way sensitivity analysis and the derivation of admissible regions. Simulation studies highlight the efficiency of our algorithm by scaling it to massive networks with up to 100’000 parameters and investigate the feasibility of generic multi-way analyses. Our routines are also showcased over two medium-sized Bayesian networks: the first modeling the country-risks of a humanitarian crisis, the second studying the relationship between the use of technology and the psychological effects of forced social isolation during the COVID-19 pandemic. An implementation of the methods using the popular machine learning library PyTorch is freely available.
\end{abstract}



\section{Introduction}
\label{sec:introduction}
Probabilistic graphical models, and specifically Bayesian networks (BNs), are a class of models that are widely used for risk assessment of complex operational systems in a variety of domains. The main reason for their success is that they provide an efficient and intuitive framework to represent the joint probability of a vector of variables of interest using a simple graph. Their use to assess the reliability of engineering, medical and ecological systems, among many others,  is becoming increasingly popular. Sensitivity analysis is a critical step for any applied real-world analysis to assess the importance of various risk factors and to evaluate the overall safety of the system under study \citep[see e.g.][for some recent examples]{goerlandt2021bayesian,makaba2021bayesian,zio2022bayesian}. 
	
As noticed by  \citet{rohmer2020uncertainties}, sensitivity analysis in BNs is usually \textit{local}, in the sense that it measures the effect of a small number of parameter variations on output probabilities of interest, while other parameters are kept fixed. In the case of a single parameter variation, sensitivity analysis is usually referred to as \textit{one-way}; otherwise, when more than one parameter is varied, it is called \textit{multi-way}. Although recently there has been an increasing interest in proposing \textit{global} sensitivity methods for BNs measuring how different factors \textit{jointly} influence some function of the model's output \citep[see e.g.][]{ballester2022computing,li2018sensitivity}, the focus of this paper still lies in local sensitivity methods.

Local sensitivity analysis in BNs can be broken down into two main steps. First, some parameters of the model are varied, and the effect of these variations on output probabilities of interest is investigated. For this purpose, a simple mathematical function, usually termed \emph{sensitivity function}, describes an output probability of interest as a function of the BN parameters \citep{castillo1997sensitivity,coupe2002properties}. Furthermore, some specific properties of such a function can be computed, for instance, the \emph{sensitivity value} or the \emph{vertex proximity}, which give an overview of how sensitive the probability of interest is to variations of the associated parameter \citep{van2007sensitivity}.  Second, once parameter variations are identified, their effect is summarized by a distance or divergence measure between the original and the varied distributions underlying the BN, most commonly the Chan-Darwiche distance \citep{chan2005distance} or the well-known Kullback-Leibler divergence. 
	
As demonstrated by \citet{uai2008}, the derivation of both the sensitivity function and its associated properties is computationally very demanding. In \citet{ballester2022you}, we introduced a novel, computationally highly-efficient method to compute all sensitivity measures of interest in one-way sensitivity analysis, which takes advantage of backpropagation and is easy to compute thanks to automatic differentiation. We now also demonstrate how the algorithm can be utilized for more generic multi-way sensitivity analyses and for deriving admissible regions \citep{uai2001}. Simulation studies show the efficiency of the approach by processing massive networks in a few seconds and demonstrate when multi-way analyses are computationally feasible. Two practical applications from real-world datasets further showcase the insights sensitivity measures can provide and the efficiency of the implemented routines. 

We have open-sourced a Python implementation using the popular machine learning library PyTorch\footnote{Available at \url{https://github.com/rballester/yodo}.}, contributing to the recent effort of promoting sensitivity analysis \citep{Douglas2020}.

\section{Bayesian networks and sensitivity analysis}
\label{sec:sens}

A BN is a probabilistic graphical model defining a factorization of the probability mass function (pmf) of a random vector using a directed acyclic graph (DAG) \citep{darwiche2009modeling,pearl}. More formally, let $[p]=\{1,\dots,p\}$ and $\bm{Y}=(Y_i)_{i\in[p]}$ be a random vector of interest with sample space $\mathbb{Y}=\times_{i\in[p]}\mathbb{Y}_i$. A BN defines the pmf $P(\bm{Y}=\bm{y})$, for $\bm{y}\in\mathbb{Y}$, as a product of simpler conditional pmfs as follows:
\begin{equation}
P(\bm{Y}=\bm{y}) = \prod_{i\in[p]}P(Y_i=y_i\;|\; \bm{Y}_{\Pi_i}=\bm{y}_{\Pi_i}),
\end{equation}
where $\bm{Y}_{\Pi_i}$ are the parents of $Y_i$ in the DAG associated to the BN.

The definition of the pmf over $\bm{Y}$, which would require defining $\#\mathbb{Y}-1$ probabilities, is thus simplified in terms of one-dimensional conditional pmfs. The coefficients of these functions are henceforth referred to as the parameters $\bm\theta$ of the model. The DAG structure may be either expert-elicited or learned from data using structural learning algorithms, and the associated parameters $\bm\theta$ can be either expert-elicited or learned using frequentist or Bayesian approaches. No matter the method used, we assume that a value for these parameters $\bm\theta$ has been chosen, which we refer to as the \textit{original value} and denote it as $\bm\theta^0$.

The DAG associated with a BN provides an intuitive overview of the relationships between variables of interest. However, it does also provide a framework to assess if any generic conditional independence holds for a specific subset of the variables via the so-called d-separation criterion \citep[see e.g.][]{pearl}. Furthermore, the DAG provides a framework for the efficient propagation of probabilities and evidence via algorithms that take advantage of the structure of the underlying DAG.

\subsection{One-way sensitivity analysis}
In practical applications, it is fundamental to extensively assess the implications of the chosen parameter values $\bm\theta^0$ to outputs of the model. In the context of BNs, this study is usually referred to as \textit{sensitivity analysis}, which can be further used during the model-building process as showcased by \citet{coupe2000sensitivity}. Let $Y_O$ be an output variable of interest and $\bm{Y}_E$ be \textit{evidential} variables, those that may be observed. The interest is in then studying how $P(Y_O=y_O\;|\; \bm{Y}_E=\bm{y}_E)$ varies when a parameter $\theta_i$ is varied. In particular, $P(Y_O=y_O\;|\; \bm{Y}_E=\bm{y}_E)$ seen as a function of $\theta_i$ is called \textit{sensitivity function} and denoted as $f(\theta_i)$.

\subsection{Proportional covariation}

Notice that when an input $\theta_i$ is varied from its original value $\theta_i^0$, the parameters from the same conditional pmf need to \textit{covary} to respect the sum-to-one condition of probabilities. When variables are binary, this is automatic since one parameter must be equal to one minus the other. However, for variables taking more than two levels, this covariation can be done in several ways \citep{renooij2014co}. We henceforth assume that whenever a parameter is varied from its original value $\theta_i^0$  to a new value $\theta_i$, then every parameter $\theta_j$ from the same conditional pmf is \textit{proportionally covaried} \citep{laskey1995sensitivity} from its original value $\theta_j^0$:

\begin{equation} \label{eq:proportional_covariation}
	\theta_j(\theta_i) = \frac{1-\theta_i}{1-\theta^0_i}\theta^0_j.
\end{equation}

Proportional covariation has been studied extensively, and its choice is motivated by a wide array of theoretical properties \citep{chan2005distance,leonelli2017sensitivity,leonelli2018geometric,renooij2014co}.

Under the assumption of proportional covariation, \citet{castillo1997sensitivity} and \citet{coupe2002properties} demonstrated that the sensitivity function is the ratio of two linear functions:
\begin{equation} \label{eq:sens}
	f(\theta_i)=\frac{c_0+c_i\theta_i}{d_0+d_i\theta_i},
\end{equation}
where $c_0,c_i,d_0,d_i\in\mathbb{R}_{+}$. \citet{van2007sensitivity}  noticed that the above expression coincides with the fragment of a rectangular hyperbola, which can be generally written as 
\begin{equation}
	f(\theta_i) = \frac{r}{\theta_i-s}+t,
\end{equation}
where
\begin{equation}
	s = -\frac{d_0}{d_i}, \;\; t=\frac{c_i}{d_i}, \;\; r = \frac{c_0}{d_i}+s t.
\end{equation}

\subsubsection{Sensitivity values}
\label{sec:sensitivity_value}

The \textit{sensitivity value}  describes the effect of infinitesimally small shifts in the parameter’s original value on the probability of interest and is defined as the absolute value of the first derivative of the sensitivity function at the original value of the parameter, i.e. $|f^{'}(\theta_i^0)|$. This can be found by simply differentiating the sensitivity function as
\begin{equation} \label{eq:first_derivative}
|f^{'}(\theta_i^0)| = \frac{|c_i d_0 - c_0 d_i|}{(d_i\theta_i^0+d_0)^2}.
\end{equation}
The higher the sensitivity value, the more sensitive the output probability to small changes in the parameter's original value. As a rule of thumb, parameters having a sensitivity value larger than one may require further investigation.

Notice that when $\bm{Y}_E$ is empty, i.e. the output probability of interest is marginal, the sensitivity function is linear in $\theta_i$. The sensitivity value is the same regardless of the original $\theta_i^0$. Therefore, in this case, the absolute value of the gradient is sufficient to quantify the effect of a parameter on an output probability of interest.

\subsubsection{Vertex proximity}
\label{sec:vertex_proximity}

\citet{van2007sensitivity} further noticed that parameters for which the sensitivity value is small may still be such that the conditional output probability of interest is very sensitive to their variations. This happens when the original parameter value is close to the \textit{vertex} of the sensitivity function, defined as the point $\theta_i^v$ at which the sensitivity value is equal to one, i.e.
\begin{equation}
	|f^{'}(\theta_i^v)| = 1.
\end{equation}
The vertex can be derived from the equation of the sensitivity function as
\begin{equation}
	\theta_i^v = \left\{
	\begin{array}{ll}
	s+\sqrt{|r|}, & \mbox{if } s <0, \\
	s - \sqrt{|r|}, & \mbox{if } s > 0.
	\end{array}
	\right.
\end{equation}
Notice that the case $s=0$ is not contemplated since it would coincide with a linear sensitivity function, not a hyperbolic one.

\textit{Vertex proximity} is defined as the absolute difference $|\theta_i^0-\theta_i^v|$. The smaller the vertex proximity, the more sensitive the output probabilities may be to parameter variations, even when the sensitivity value is small. 

\subsubsection{Other metrics}

\label{sec:other_metrics}

Given the coefficients $c_0, c_i,d_0, d_i$ of Equation~(\ref{eq:sens}), it is straightforward to derive any property of the sensitivity function besides the sensitivity value and the vertex proximity. Here we propose the use of two additional metrics. The first is the absolute value of the second derivative of the sensitivity function at the original parameter value, which can be easily computed as: 
\begin{equation} \label{eq:second_derivative}
	|f''(\theta_i^0)| = \frac{2 d_{i} \left|c_{i} d_{0} - c_{0} d_{i}\right|}{\left(d_{i} \theta_i^0 + d_{0}\right)^{3}}.
\end{equation}
Similarly to the sensitivity value, high values of the second derivative at $\theta_i^0$ indicate parameters that could highly impact the probability of interest. 

The second measure  is the maximum of the first derivative of the sensitivity function over the interval $[0, 1]$ in absolute value, which we find easily by noting that the denominator of Equation~(\ref{eq:first_derivative}) is a parabola:
\begin{equation}
	\max_{\theta_i \in [0, 1]}|f'(\theta_i)| = \begin{cases}
		\infty & \mbox{ if } -d_0 / d_i \in [0, 1] \\
		\max \{|c_i d_0 - c_0 d_i| / d_0^2, |c_i d_0 - c_0 d_i| / (d_i + d_0)^2\} & \mbox{ otherwise.} \\
	\end{cases}
\end{equation}
Again high values indicate parameters whose variations can lead to a significant change in the output probability of interest.

\subsection{Multi-way sensitivity analysis}

In many practical applications, there is interest in assessing the effect of simultaneous variations of multiple parameters on the output of interest. This is called a \textit{multi-way sensitivity analysis}. Although there have been some attempts to study the theoretical properties and computational efficiency of these more generic analyses \citep[see e.g.][]{bolt2014local,uai2004,uai2000,leonelli2017sensitivity,leonelli2018geometric}, in practice, they are not as common as one-way analyses.

\subsubsection{General formulation}

Suppose now that $n$ parameters $\bm{\theta}_n=(\theta_1,\dots,\theta_n)$ are simultaneously varied. By default, these parameters are taken from different conditional pmfs so that they are independent of each other \citep{van2007sensitivity}. In the binary case, this is natural since only one parameter per pmf can be varied since the other is functionally related. The other parameters from conditional pmfs including  $\theta_1,\dots,\theta_n$, are proportionally covaried, as for the one-way analysis \citep[see][for a formal discussion]{leonelli2018geometric}. The effect of varying the parameters $\bm{\theta}_n$ on a probability of interest $P(Y_O=y_O\;|\; \bm{Y}_E=\bm{y}_E)$ is captured by the n-way sensitivity function, which is equal to

\begin{equation}
f(\bm{\theta}_n)=\frac{\sum_{K\in\mathcal{P}([n])}c_K\prod_{i\in K}\theta_i}{\sum_{K\in\mathcal{P}([n])}d_K\prod_{i\in K}\theta_i},
\end{equation}
where $\mathcal{P}$ denotes the power set and $c_K,d_K\in\mathbb{R}$, $K\in\mathcal{P}([n])$, are constants computed from the non-varied parameters. For instance, a 2-way sensitivity function can be written as:

\begin{equation}
f(\theta_1,\theta_2)=\frac{c_0+c_1\theta_1+c_2\theta_2+c_{12}\theta_1\theta_2}{d_0+d_1\theta_1+d_2\theta_2+d_{12}\theta_1\theta_2}
\label{eq:2way_sensitivity}
\end{equation}

An n-way sensitivity function, in general, requires the computation of $2^{n+1}$ constants and is thus computationally expensive. Furthermore, the number of combinations of parameters for which the sensitivity function has to be constructed increases: see Section \ref{sec:yodomulti} for a discussion.

\subsubsection{Maximum n-way sensitivity values}

While for one-way sensitivity analysis, one can uniquely talk about the derivative of the sensitivity function, for multi-valued functions, there are multiple directions at which the derivative could be computed, as noted by \citep{bolt2014local}, and hence the notion of \textit{directional derivative}. However, basic calculus tells us that the maximum directional derivative of a function $f$ at a point $\bm{\theta}_n$ equals the length of the gradient vector at $\bm{\theta}_n$, i.e. $|\Delta f(\bm{\theta}_n)|$. This observation led to the definition of the sensitivity value for an n-way sensitivity function as the maximum one out of all possible directional derivatives \citep{bolt2014local}. For a vector of parameters $\bm{\theta}_n$ with original values $\bm{\theta}_n^0$ the \textit{maximum n-way sensitivity value} is defined as
\begin{equation}
\label{eq:max}
sv_{\text{max}}^{\bm{\theta}_n}=|\Delta f(\bm{\theta}_n^0)|,
\end{equation}
where $f$ is the associated n-way sensitivity function.

By definition, the maximum n-way sensitivity value would first require the derivation of the n-way sensitivity function and, subsequently, the computation of its gradient. As noticed already, this direct approach would be computationally too expensive. However, \citet{bolt2014local} demonstrated that $sv_{\text{max}}^{\bm{\theta}_n}$ could be easily computed from the sensitivity values of one-way sensitivity functions. Let $c_0^i,c_i,d_0^i,d_i$ be the coefficients of the one-way sensitivity function for the variation of the parameter $\theta_i$ in $\bm{\theta}_n$. Then:
\begin{equation}
\label{eq:max2}
sv_{\text{max}}^{\bm{\theta}_n}=\frac{1}{P(\bm{Y}_E=\bm{y}_E)^2}\sqrt{\sum_{i\in[n]} (c_id_0^i-c_0^id_i)^2}.
\end{equation}

Therefore if an efficient method for computing the coefficients of one-way sensitivity functions exists, then maximum n-way sensitivity values can be equally efficiently derived.

\subsection{Admissible regions}

In many applied situations, the object of interest is not a probability per se, but rather the most likely value of a variable, possibly conditional on a specific subset of evidence. This is the case for classification problems where a Bayes classifier is used: an unlabeled observation exhibiting a specific evidence pattern is classified according to the most likely value. BNs designed explicitly for this task are usually called Bayesian network classifiers \citep{bielza2014discrete,friedman1997bayesian}.  

Although sensitivity methods for this type of classification problem have been discussed \citep{bolt2017balanced}, sensitivity values and related measures are often not particularly useful. \citet{uai2001} demonstrated that parameters with a small sensitivity value might induce a change in the classification rule, or equally in the most likely value, for just a slight deviation from its original value. For this reason, they introduced the concept of \emph{admissible region}, which captures the extent to which a parameter can be varied without inducing a change in the most likely value for the variable of interest.

For ease of notation,  we consider here a variable of interest $Y_O$ taking two possible levels $y_O$ and $y_O'$ (thus, we consider the most common binary classification problem). Consider also possible evidence $\bm{Y}_E=\bm{y}_E$, a perturbed parameter $\theta_i$ and suppose that $P(Y_O=y_O|\bm{Y}_E=\bm{y}_E)>P(Y_O=y_O'|\bm{Y}_E=\bm{y}_E)$, without loss of generality. The admissible region $R_i$ is formally defined as the interval of values for $\theta_i$
\begin{equation}
(\max\{\theta_i^0-r,0\},\min\{\theta_i^0+s,1\}), \quad \quad r,s,\in\mathbb{R},
\end{equation} 
for which $P(Y_O=y_O|\bm{Y}_E=\bm{y}_E)>P(Y_O=y_O'|\bm{Y}_E=\bm{y}_E)$. The wider the interval $R_i$, the less influential the parameter is for the most likely value.

\citet{uai2001} and \citet{van2007sensitivity} already demonstrated that such regions could be computed from the one-way sensitivity functions by identifying the points at which the sensitivity functions intersect. However, they did not explicitly write the admissible regions as a function of the sensitivity functions' coefficients to our knowledge. Let $c_0,c_i,d_0,d_i$ be the coefficients of the sensitivity function for the event $Y_O=y_O|\bm{Y}_E=\bm{y}_E$. It follows that the sensitivity function for $Y_O=y_O'|\bm{Y}_E=\bm{y}_E$ must be equal to
\begin{equation}
\frac{(d_0-c_0)+(d_i-c_i)\theta_i}{d_0+d_i\theta_i}.
\end{equation}
By equating the two sensitivity functions, we find that 
\begin{equation}
\label{eq:ad1}
R_i = \left\{
\begin{array}{ll}
\left(0,\min\{\frac{d_0-2c_0}{2c_i-d_i},1\}\right) & \mbox{if } \theta_i^0\leq\frac{d_0-2c_0}{2c_i-d_i}\\
\left(\max\{0,\frac{d_0-2c_0}{2c_i-d_i}\},1\right) & \mbox{otherwise}
\end{array}
\right.
\end{equation}

In the case of $E=\emptyset$, i.e. no evidence, the expression for the admissible regions simplifies to:

\begin{equation}
\label{eq:ad2}
R_i = \left\{
\begin{array}{ll}
\left(0,\min\{\frac{1-2c_0}{2c_i},1\}\right) & \mbox{if } \theta_i^0\leq \frac{1-2c_0}{2c_i}\\
\left(\max\{0,\frac{1-2c_0}{2c_i}\},1\right) & \mbox{otherwise}
\end{array}
\right.
\end{equation}

Therefore, given an efficient method to compute one-way sensitivity functions, admissible regions for all individual parameters can be equally efficiently derived.

\section{The YODO method}

The YODO (You Only Derive Once) method was first introduced in \citet{ballester2022you} to compute the one-way sensitivity measures discussed in Sections \ref{sec:sensitivity_value}-\ref{sec:other_metrics}. We first review it and then discuss its use in multi-way sensitivity analysis.

\subsection{YODO for one-way sensitivity analysis}
\subsubsection{First case: Marginal probability as a function of interest}
\label{sec:marginal_case}

Suppose $f(\theta_i) = P(Y_O = y_O) = c_0+c_i \theta_i $ assuming proportional covariation as $\theta_i$ varies. Let $\theta_{j_1}, \dots, \theta_{j_n}$ be the other parameters of the same conditional pmf as $\theta_i$, i.e. they are all bound by the sum-to-one constraint $\theta_i + \theta_{j_1} + \dots + \theta_{j_n} = 1$.
First, we rewrite $f$ as
\begin{equation}
	f(\theta_i) = g(\theta_i, \theta_{j_1}(\theta_i), \dots, \theta_{j_n}(\theta_i))
\end{equation}
and we show how to obtain $f'(\theta_i)$ provided that we can compute the gradient $\nabla g$ with respect to symbols $\theta_i, \theta_{j_1}, \dots, \theta_{j_n}$ (see Section~\ref{sec:computing_the_gradient} for details on the latter).

By the generalized chain rule, it holds that
\begin{equation} \label{eq:generalized_chain_rule}
	f'(\theta_i) = \frac{\partial g}{\partial \theta_i} \cdot 1 + \frac{\partial g}{\partial \theta_{j_1}} \cdot \frac{d\theta_{j_1}}{d\theta_i} + \dots + \frac{\partial g}{\partial \theta_{j_n}} \cdot \frac{d \theta_{j_n}}{d\theta_i}.
\end{equation}

By deriving Equation~(\ref{eq:proportional_covariation}), we have that for all $1 \le m \le n$:

\begin{equation} \label{eq:derivative}
	\frac{d\theta_{j_m}}{d\theta_i} = \frac{-\theta_{j_m}^0}{1 - \theta_i^0}
\end{equation}

%
%

and, therefore,
\begin{equation}
	f'(\theta_i) = \frac{\partial g}{\partial \theta_i} - \frac{(\partial g / \partial \theta_{j_1}) \cdot \theta_{j_1}^0 + \dots + (\partial g / \partial \theta_{j_n}) \cdot \theta_{j_n}^0}{1 - \theta_i^0}.
\end{equation}

Last, since $f(\theta_i) = P(\bm{Y}_O = \bm{y}_O) = c_0+c_i \theta_i $, we easily find the parameters $c_0, c_i$:

\begin{equation}
	\begin{cases}
		c_i = f'(\theta_i^0) \\
		c_0 = P(\bm{Y}_O = \bm{y}_O) - c_i \theta_i^0.
	\end{cases}
\end{equation}

\subsubsection{Second case: Conditional probability as a function of interest}

When $f(\theta_i) = P(Y_O=y_O\;|\; \bm{Y}_E=\bm{y}_E) = P(Y_O=y_O, \bm{Y}_E=\bm{y}_E) / P(\bm{Y}_E = \bm{y}_E)$, we simply repeat the procedure from Sec.~\ref{sec:marginal_case} twice:

\begin{enumerate}
	\item We first apply it to $P(Y_O=y_O, \bm{Y}_E=\bm{y}_E)$ to obtain $c_0$ and $c_i$;
	\item we then apply it to $P(\bm{Y}_E=\bm{y}_E)$ to obtain $d_0$ and $d_i$.
\end{enumerate}

\subsubsection{Computing the gradient $\nabla g$}
\label{sec:computing_the_gradient}


Let $\bm{Y}_K = \bm{y}_K$ be a subset of the network variables taking some evidence values (this could be $K = O$ or $K = O \cup E$; hence we cover the two cases above).

We start by moralizing the BN into a Markov random field (MRF) $\mathcal{M}$. This marries all variable parents together and, for each conditional probability table (now called \emph{potential}), drops the sum-to-one constraint; see e.g.~\citep{Darwiche2009} for more details. Next, we impose the evidence $\bm{Y}_K = \bm{y}_K$ by defining $\mathcal{M}^{\bm{Y}_K = \bm{y}_K}$ as a new MRF that results from substituting each potential $\Phi_{i_1, \dots, i_M}(x_{i_1}, \dots, x_{i_M})$ by a new potential $\widehat{\Phi}_{i_1, \dots, i_M}$ defined as follows:

\begin{equation}
\begin{split}\widehat{\Phi}_{i_1, \dots, i_M}(Y_{i_1} = x_{i_1}, \dots, Y_{i_M} = x_{i_M}) = \\
\begin{cases}
0 & \mbox{if } \exists m, k \; \vert \; i_m = k \wedge x_{i_m} \ne y_{i_m} \\
\Phi_{i_1, \dots, i_M}(Y_{i_1} = x_{i_1}, \dots, Y_{i_M} = x_{i_M}) & \mbox{otherwise} \\
\end{cases}
\end{split}
\end{equation}

In other words, we copy the original potential but zero-out all entries that do not honor the assignment of values $\bm{Y}_K = \bm{y}_K$. See Table~\ref{tab:potential_example} for an example using a bivariate potential.
\begin{table}
\begin{center}
	\begin{subtable}[h]{0.48\textwidth}
		\begin{tabular}{c | c | c | c} 
			& $Y_2 = 1$ & $Y_2 = 2$ & $Y_2 = 3$ \\ [0.5ex] 
			\hline
			$Y_1 = 1$ & 0.8 & 0.1 & 0.1 \\ 
			\hline
			$Y_1 = 2$ & 0.3 & 0.5 & 0.2 \\
			\hline
			$Y_1 = 3$ & 0.1 & 0.2 & 0.7 \\
			\hline
		\end{tabular}
		\caption{$\Phi_{1, 2}(y_1, y_2)$}
	\end{subtable}
	\begin{subtable}[h]{0.48\textwidth}
		\begin{tabular}{c | c | c | c} 
			& $Y_2 = 1$ & $Y_2 = 2$ & $Y_2 = 3$ \\ [0.5ex] 
			\hline
			$Y_1 = 1$ & 0.0 & 0.0 & 0.1 \\ 
			\hline
			$Y_1 = 2$ & 0.0 & 0.0 & 0.2 \\
			\hline
			$Y_1 = 3$ & 0.0 & 0.0 & 0.7 \\
			\hline
		\end{tabular}
		\caption{$\widehat{\Phi}_{1, 2}(y_1, y_2)$}
	\end{subtable}
 \end{center}
	\caption{Left: example potential of an MRF $\mathcal{M}$ for variables $Y_1$ and $Y_2$, each with three levels $\{1, 2, 3\}$. Right: corresponding potential for $\mathcal{M}^{Y_2 = 3}$.}
	\label{tab:potential_example}
\end{table}

Intuitively, the modified MRF $\mathcal{M}^{\bm{Y}_K = \bm{y}_K}$ represents the unnormalized probability for all variable assignments that are compatible with $\bm{Y}_K = \bm{y}_K$. In particular, if $\mathcal{M}_{\bm{Y}_K}$ denotes the marginalization of a network $\mathcal{M}$ over all variables in $\bm{Y}_K$, we have that $(\mathcal{M}^{\bm{Y}_K = \bm{y}_K})_{\bm{Y}} = P(\bm{Y}_K = \bm{y}_K)$. In other words, computing $g$ reduces to marginalizing our MRF. In this paper, we marginalize it exactly using the variable elimination (VE) algorithm \citep[see e.g][]{Darwiche2009}. This method is differentiable w.r.t. all parameters $\bm{\theta}$ since VE only relies on variable summation and factor multiplication. Any other differentiable inference algorithm could be used as well \citep[for instance, the junction tree algorithm as in][]{uai2000}. This step, evaluating the function $g$, is known as the \emph{forward pass} in the neural network literature. Next, we backpropagate the previous operation (a step known as the \emph{backward pass}) to build the gradient $\nabla g$. Crucially, note that backpropagation yields $\partial g / \partial \theta$ for every parameter $\theta \in \bm{\theta}$ of the network at once, not just an individual $\theta_i$. Last, we obtain parameters $c_0,c_i,d_0,d_i$ as detailed before, and use them to compute the metrics of Sections~\ref{sec:sensitivity_value}-\ref{sec:other_metrics} for each $\theta_i$.

Note the advantages of this approach as compared to other alternatives. For example, symbolically deriving the gradient of $g$ would be cumbersome and depend on the target network topology and definition of the probability of interest \citep{darwiche2003differential}. Automatic differentiation avoids this by evaluating the gradient numerically using the chain rule. Furthermore, finding the gradient using finite differences would require evaluating $g$ twice per parameter $\theta_i$. In contrast, automatic differentiation only requires a forward and backward pass to find the entire gradient --in our experiments, roughly the time of just two marginalization operations (see below).


\subsubsection{Additional one-way information}
Although YODO is specifically designed to compute the coefficients of the one-way sensitivity function of Equation (\ref{eq:sens}), it further provides all the information to answer additional sensitivity questions:
\begin{itemize}
\item It provides the admissible regions for every parameter in $\bm{\theta}$ concerning the event of interest $Y_O=y_O|\bm{Y}_E=\bm{y}_E$, since they formally only depend on the coefficients $c_0,c_i,d_0,d_i$ as shown in Equations (\ref{eq:ad1}) and (\ref{eq:ad2}).
\item  It can quickly find the parameters that do not affect the output probability of interest. This set is usually called the \textit{parameter sensitivity set} \citep{coupe2002properties}.  This consists of the parameters $\theta_i$ for which $c_i$ and/or $d_i$ are non-zero.
\item It identifies whether a parameter change leads to a monotonically increasing or decreasing sensitivity function, as already addressed in \citet{bolt2017structure}. Again this can be straightforwardly derived by checking the sign of $c_id_0-c_0d_i$: see Equation (\ref{eq:first_derivative}).
\end{itemize}

\subsection{YODO for multi-way sensitivity analysis}
\label{sec:yodomulti}

Although there would be no difficulty in conceptually considering simultaneous variations of multiple parameters, we restrict our attention to 2-way sensitivity analyses where only pairs of parameters are varied. This is because: (i) sensitivity functions cannot be visualized in higher dimensions; (ii) the number of groups of parameters grows exponentially; (iii) most critically, the associated measures are challenging to interpret, similar to higher-order interactions in standard statistical models  \citep[see e.g.][]{hayes2012cautions}.

The 2-way version of the sensitivity function considered before would entail computing the unknowns $c_{12}$ and $d_{12}$ from Equation~\ref{eq:2way_sensitivity}. This can be achieved by computing the Hessian, rather than the gradient, in the previous calculations, which is supported in most modern autodifferentiation packages. However, the sheer size of the Hessian (up to $10^{10}$ in the networks considered in Sec.~\ref{sec:simulation_study}) would make the interpretation of such indices a challenge of its own.


Therefore, we advocate that the maximum n-way sensitivity value is the most valuable and versatile tool for multi-way sensitivity analysis. From its definition in Equation (\ref{eq:max}), it is clear that it can be instantaneously computed for a specific combination of parameters $\bm{\theta}_n$ once the YODO algorithm has been run. Still, even when focusing on $n=2$, the possible $\bm{\theta}_n$ can become overwhelmingly large for medium-sized BNs. To address this, we introduce an algorithm to obtain the top $K$ $sv_{\text{max}}$ pairs efficiently by noting that parameters $\bm{\theta}$ contribute to Equation~(\ref{eq:max2}) independently from each other. We use a priority queue and proceed in a dynamic programming fashion, whereby we start with a pool $\mathcal{P}$ of $K$ best candidates and keep track of $\max_j sv_{\text{max}}^{\theta_i, \theta_j}$ for all $i \in \mathcal{P}$. The top $K$ pairs are guaranteed to be found after $K$ steps. The algorithm relies on sorting $n$ elements and on $K$ insertions and deletions on the queue and runs in $O(n \log n + K^2 \log K)$ operations. See Algorithm~\ref{alg:top_svmax_pairs} for all details.

\begin{algorithm}
	\footnotesize
	\caption{Algorithm to find the top $K$ maximum 2-way sensitivity values $sv_{\text{max}}^{\theta_i, \theta_j}$ for any $i, j \in [n]$.}
	\begin{algorithmic}[1]
		\STATE // Gather contributions to Eq.~\ref{eq:max2} from every BN parameter $\theta_i$
		\STATE $v \gets $ empty vector
		\FOR{$i \gets 1$ to $n$}
			\STATE $v_i \gets (c_id_0^i-c_0^id_i)^2$
		\ENDFOR
		\STATE $v \gets \text{sortDescending}(v)$
		\STATE
		\STATE // Populate the queue with $K$ initial candidates
		\STATE $q \gets $ empty priority queue
		\FOR{$i \gets 1$ to $K$}
			\STATE $q.\text{put}(\frac{1}{P(\bm{Y}_E=\bm{y}_E)^2}\sqrt{v_i + v_{i+1}}, i, i+1)$ // First element acts as queue's key
		\ENDFOR

		\STATE
		\STATE // Read the queue's largest $K$ keys while updating it
		\STATE $w \gets $ empty vector
		\FOR{$k \gets 1$ to $K$}
			\STATE $(v, i, j) \gets q.\text{get()}$
			\STATE $w_k \gets v$
			\IF{$j < n$}
				\STATE // Insert next pair candidate
				\STATE $q.\text{put}(\frac{1}{P(\bm{Y}_E=\bm{y}_E)^2}\sqrt{v_i + v_{j+1}}, i, j+1)$
			\ENDIF
		\ENDFOR
		\RETURN $w$
	\end{algorithmic}
	\label{alg:top_svmax_pairs}
\end{algorithm}

\subsection{Implementation}
In order to perform variable elimination efficiently, we note that the problem of graphical model marginalization is equivalent to that of tensor network contraction~\citep{RS:18}, and use the library \emph{opt\_einsum}~\citep{SG:18} which offers optimized heuristics for the latter. As backend, we use the state-of-the-art machine learning library \emph{PyTorch}~\citep{PGM+:19}, version 1.13.1, to do all operations between tensors and then perform backpropagation on them. We use \emph{pgmpy}~\citep{ankan2015pgmpy} for reading and moralizing BNs.  

\section{Results}
\label{sec:appl}

We first study the method's scalability by testing it on large networks with hundreds of nodes and arcs and up to $10^5$ parameters; we then overview the insights revealed by our method when applied to two Bayesian networks. All experiments were run on a 4-core i5-6600 3.3GHz Intel workstation with 16GB RAM.

\subsection{Simulation study}  \label{sec:simulation_study}

First, we run our method over the 10 Bayesian networks considered in \citet{scutari2019learns}. As a baseline, we use the numerical estimation of each sensitivity value via finite differences, whereby we slightly perturb each parameter $\theta_i$ and measure the impact on $f$. As a probability of interest, we set $P(A = a | B = b)$, where $A, B, a, b$ were two variables, and two levels picked randomly, respectively. Each timing is the average of three independent runs. Results are reported in Table~\ref{tab:many_network_study}, which shows that YODO outperforms the baseline by several orders of magnitude and that computing the most relevant 2-way sensitivity values takes in the order of 2s at most. 

\begin{table} \small
\centering
\scalebox{0.75}{
	\begin{tabular}{lrrrrrrr}
\toprule
{} &  \#nodes &  \#arcs &  \#parameters &  Treewidth &  Time (fin. diff.) &  Time (autodiff.) &  Time (svmax) \\
Network    &          &         &               &            &                    &                   &               \\
\midrule
child      &       20 &      30 &           344 &          3 &           5.901379 &          0.026062 &      0.011229 \\
water      &       32 &     123 &         13484 &         10 &         246.183577 &          0.057886 &      0.242638 \\
alarm      &       37 &      65 &           752 &          4 &          12.021088 &          0.039274 &      0.019980 \\
hailfinder &       56 &      99 &          3741 &          4 &          58.217657 &          0.062527 &      0.092034 \\
hepar2     &       70 &     158 &          2139 &          6 &         100.047695 &          0.089120 &      0.052889 \\
win95pts   &       76 &     225 &          1148 &          8 &          38.573179 &          0.092268 &      0.031935 \\
pathfinder &      109 &     208 &         97851 &          6 &        9254.500155 &          0.182588 &      1.848624 \\
munin1     &      186 &     354 &         19226 &         11 &      148307.455340 &         16.085417 &      1.194862 \\
andes      &      223 &     626 &          2314 &         17 &         238.634045 &          0.311464 &      0.070569 \\
pigs       &      441 &     806 &          8427 &         10 &        1544.150893 &          0.568345 &      0.213123 \\
\bottomrule
\end{tabular}

 }
	\caption{Our method was applied to 10 Bayesian networks, here sorted by the number of nodes. All times are in seconds. The times for the baseline (third-to-last column) were estimated as the total number of parameters in the network and the time needed to estimate one sensitivity value numerically. Treewidths were found with the \emph{NetworkX} graph library~\cite{HSS:08}. The last column reports the time needed to find the top 20 $sv_{\mbox{max}}$ pairs based on existing YODO gradients.}
	\label{tab:many_network_study}
\end{table}

\subsection{Risk assessment for humanitarian crises and disasters}

\begin{table} \small
	\centering
 \scalebox{0.65}{
	\begin{tabular}{lrrr}
\toprule
 Variable &  Abbreviation &  Risk Dimension &  Category   \\
\midrule
\textbf{\cellcolor{gray!2}{Earthquake}}   &   \cellcolor{gray!2}{EARTHQUAKE}   &     \cellcolor{gray!2}{Hazard and Exposure} & \cellcolor{gray!2}{Natural} \\
\textbf{\cellcolor{gray!2}{Tsunami}}  &   \cellcolor{gray!2}{TSUNAMI}   &     \cellcolor{gray!2}{Hazard and Exposure} & \cellcolor{gray!2}{Natural} \\
\textbf{\cellcolor{gray!2}{Flood}} & \cellcolor{gray!2}{FLOOD} &  \cellcolor{gray!2}{Hazard and Exposure} & \cellcolor{gray!2}{Natural} \\
\textbf{\cellcolor{gray!2}{Tropical Cyclone}} & \cellcolor{gray!2}{TROP\_CYC} &  \cellcolor{gray!2}{Hazard and Exposure} & \cellcolor{gray!2}{Natural} \\
\textbf{\cellcolor{gray!2}{Drought}} & \cellcolor{gray!2}{DROUGHT} &\cellcolor{gray!2}{Hazard and Exposure} & \cellcolor{gray!2}{Natural} \\
\textbf{\cellcolor{gray!2}{Epidemic}} & \cellcolor{gray!2}{EPIDEMIC} & \cellcolor{gray!2}{Hazard and Exposure} & \cellcolor{gray!2}{Natural}\\
\textbf{\cellcolor{gray!3}{Projected Conflict Risk}} & \cellcolor{gray!3}{PCR} & \cellcolor{gray!3}{Hazard and Exposure} & \cellcolor{gray!3}{Human} \\
\textbf{\cellcolor{gray!3}{Current Highly Violent Conflict Intensity}} & \cellcolor{gray!3}{CHVCI} & \cellcolor{gray!3}{Hazard and Exposure} & \cellcolor{gray!3}{Human} \\
\textbf{\cellcolor{gray!5}{Development and Deprivation}} & \cellcolor{gray!5}{D\_AND\_D} & \cellcolor{gray!5}{Vulnerability} & \cellcolor{gray!5}{Socio-Economic}\\
\textbf{\cellcolor{gray!5}{Economic Dependency}} & \cellcolor{gray!5}{ECON\_DEP} &  \cellcolor{gray!5}{Vulnerability} & \cellcolor{gray!5}{Socio-Economic}\\
\textbf{\cellcolor{gray!6}{Unprotected People}} & \cellcolor{gray!6}{UNP\_PEOPLE} & \cellcolor{gray!6}{Vulnerability} & \cellcolor{gray!6}{Vulnerable Groups}\\
\textbf{\cellcolor{gray!6}{Other Vulnerable Groups}} & \cellcolor{gray!6}{OTHER\_VULN\_GROUPS} & \cellcolor{gray!6}{Vulnerability} & \cellcolor{gray!6}{Vulnerable Groups}\\
\textbf{\cellcolor{gray!6}{Children U5}} & \cellcolor{gray!6}{CHILDREN\_U5} & \cellcolor{gray!6}{Vulnerability} & \cellcolor{gray!6}{Vulnerable Groups}\\
\textbf{\cellcolor{gray!6}{Food Security}} & \cellcolor{gray!6}{FOOD\_SEC} &\cellcolor{gray!6}{Vulnerability} & \cellcolor{gray!6}{Vulnerable Groups}\\
\textbf{\cellcolor{gray!6}{Recent Shocks}} & \cellcolor{gray!6}{RECENT\_SHOCKS} & \cellcolor{gray!6}{Vulnerability} & \cellcolor{gray!6}{Vulnerable Groups}\\
\textbf{\cellcolor{gray!6}{Health Conditions}} & \cellcolor{gray!6}{HEALTH\_COND} & \cellcolor{gray!6}{Vulnerability} & \cellcolor{gray!6}{Vulnerable Groups}\\
\textbf{\cellcolor{gray!8}{Governance}} & \cellcolor{gray!8}{GOVERNANCE} & \cellcolor{gray!8}{Lack of Coping Capacity} & \cellcolor{gray!8}{Institutional} \\
\textbf{\cellcolor{gray!9}{Communication}} & \cellcolor{gray!9}{COMMUNICATION} & \cellcolor{gray!9}{Lack of Coping Capacity} & \cellcolor{gray!9}{Infrastructure} \\
\textbf{\cellcolor{gray!9}{Physical Infrastructure}} & \cellcolor{gray!9}{PHYS\_INFRA} & \cellcolor{gray!9}{Lack of Coping} \cellcolor{gray!9}{Capacity} & \cellcolor{gray!9}{Infrastructure} \\
\textbf{\cellcolor{gray!9}{Access to Health System}} & \cellcolor{gray!9}{ACCESS\_TO\_HEALTH} &\cellcolor{gray!9}{Lack of Coping Capacity} & \cellcolor{gray!9}{Infrastructure} \\ 
\bottomrule
\end{tabular}}
	\caption{Variables considered for the humanitarian network from the \citet{INFORM} dataset.
}
	\label{table:var1}
\end{table}

We next extend the analysis of \citet{ballester2022you}, which only focused on one-way indices, to assess the country-level risk associated with humanitarian crises and disasters. The data was collected from INFORM \citep{INFORM} and consists of 20 drivers of disaster risk covering natural, human, socio-economic, institutional, and infrastructure factors that influence the country-level risk of a disaster, together with a final country risk index which summarizes how exposed a country is to the possibility of a humanitarian disaster. Table \ref{table:var1} reports an overview of the twenty drivers considered, which cover three main risk dimensions: Hazard and exposure (natural/human); Vulnerability (Socio-economic/Vulnerable groups); Lack of coping capacity (institutional/infrastructure). All variables take values between zero and ten. Using the equal-length method, they have been discretized into three categories (low/0, medium/1, high/2). The dataset comprises 190 countries.

\begin{figure}
    \centering
    \includegraphics[scale=0.6]{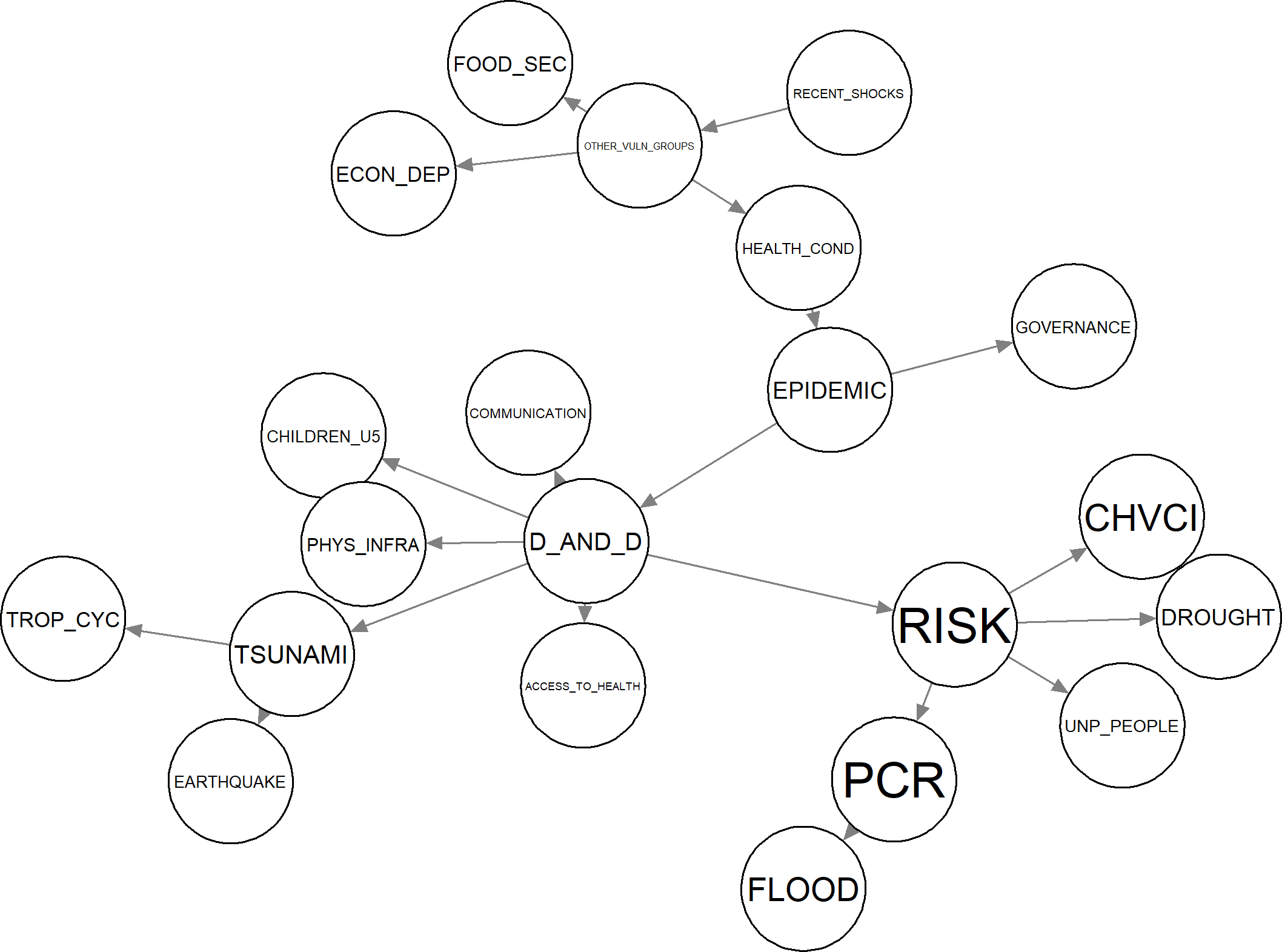}
    \caption{BN learned over the \citet{INFORM} dataset for country-level disaster risk.}
    \label{fig:bn1}
\end{figure}

Similar to \citet{qazi2021assessment}, a BN is learned using the \texttt{hc} function of the \texttt{bnlearn} package and is reported in Figure \ref{fig:bn1}. A complete interpretation of the learned DAG is beyond the scope of this paper. However, it can be noticed that most risk factors are independent of the overall country-risk given the development and deprivation index (D AND D).

\begin{table} \tiny
	\centering
 \scalebox{0.9}{
	\begin{tabular}{llllll}
\toprule
Parameter &    Value & \specialcell{Sensitivity \\ value $\downarrow$} & \specialcell{Proxi-\\mity} & $2^{\mbox{nd}}$ deriv. &
\specialcell{Largest \\ $1^{\mbox{st}}$ deriv.} \\
\midrule
RISK = high $\vert$ D\_AND\_D = low                    &               $0.0012$ 
&           $0.914$ &   $0.056$ &          $1.437$ &                  $0.916$ 
\\
FLOOD = high $\vert$ PCR = low                       &                $0.107$ & 
$0.722$ &  $0.0534$ &           $4.059$ &                  $1.475$ \\
FLOOD = medium $\vert$ PCR = low                       &                $0.469$ 
&           $0.645$ &   $0.718$ &           $3.238$ &                 $\infty$ 
\\
FLOOD = low $\vert$ PCR = low                       &                $0.425$ &  
$0.645$ &   $0.718$ &           $3.238$ &                 $\infty$ \\
RISK = high $\vert$ D\_AND\_D = high                    &                 $0.34$
&           $0.555$ &   $0.731$ &          $0.387$ &                  $0.714$ 
\\
RISK = high $\vert$ D\_AND\_D = medium                    &               
$0.0686$ &           $0.467$ &   $1.002$ &          $0.295$ &                  
$0.488$ \\
EPIDEMIC = high $\vert$ HEALTH\_COND = low            &                $0.148$ &
$0.295$ &   $1.167$ &          $0.231$ &                  $0.332$ \\
D\_AND\_D = high $\vert$ EPIDEMIC = medium                &               
$0.0742$ &           $0.238$ &   $1.834$ &          $0.133$ &                  
$0.249$ \\
PCR = high $\vert$ RISK = medium                        &                $0.278$
&           $0.226$ &     $0.6$ &           $0.395$ &                  $0.394$ 
\\
PCR = high $\vert$ RISK = low                        &               $0.0266$ & 
$0.204$ &   $0.694$ &           $0.322$ &                  $0.213$ \\
FLOOD = high $\vert$ PCR = high                       &                $0.509$ &
$0.196$ &   $0.475$ &           $0.46$ &                  $1.211$ \\
FLOOD = high $\vert$ PCR = medium                       &                $0.136$
&           $0.167$ &   $0.787$ &            $0.25$ &                  $0.206$ 
\\
D\_AND\_D = high $\vert$ EPIDEMIC = high                &                $0.787$
&           $0.159$ &   $4.159$ &         $0.0459$ &                  $0.202$ 
\\
D\_AND\_D = high $\vert$ EPIDEMIC = low                &               $0.0411$ 
&           $0.153$ &   $2.984$ &         $0.0625$ &                  $0.156$ 
\\
RISK = low $\vert$ D\_AND\_D = high                    &               $0.0208$ 
&           $0.151$ &   $2.319$ &         $0.0796$ &                  $0.274$ 
\\
HEALTH\_COND = medium $\vert$ OTHER\_VULN\_GROUPS = low   &                 
$0.05$ &           $0.151$ &   $3.026$ &          $0.061$ &                  
$0.154$ \\
HEALTH\_COND = low $\vert$ OTHER\_VULN\_GROUPS = low   &                $0.949$ 
&            $0.15$ &   $3.036$ &         $0.0606$ &                  $0.154$ 
\\
PCR = low $\vert$ RISK = high                        &              $0.00521$ & 
$0.15$ &   $3.023$ &         $0.0609$ &                  $0.236$ \\
D\_AND\_D = medium $\vert$ EPIDEMIC = high                &                
$0.176$ &           $0.148$ &   $5.393$ &         $0.0338$ &                   
$0.18$ \\
PCR = high $\vert$ RISK = high                        &                $0.943$ &
$0.148$ &   $3.092$ &         $0.0588$ &                  $0.224$ \\
\bottomrule
\end{tabular}

}
	\caption{Four sensitivity metrics for the top 20 parameters of the humanitarian crisis network, when the probability of interest is $P(\mbox{RISK = high} | \mbox{FLOOD = high})$.
}
	\label{tab:many_parameter_study}
\end{table}

As an illustration of the YODO method, we compute here all sensitivity measures for the conditional probability of a high risk of disaster (RISK = 2) conditional on a high risk of flooding (FLOOD = 2). Computing all metrics for all 183 network parameters with our method took only 0.055 seconds. The results are reported in Table \ref{tab:many_parameter_study} for the 20 most influential parameters according to the sensitivity value. It can be noticed that the most influential parameters come from the conditional distributions of the overall risk given the development and deprivation index (D AND D), as well as from the conditional distribution of the flooding index given a projected conflict risk index (PCR) equal to low. 

\begin{figure}
	\begin{center}
		\includegraphics[scale = 0.4]{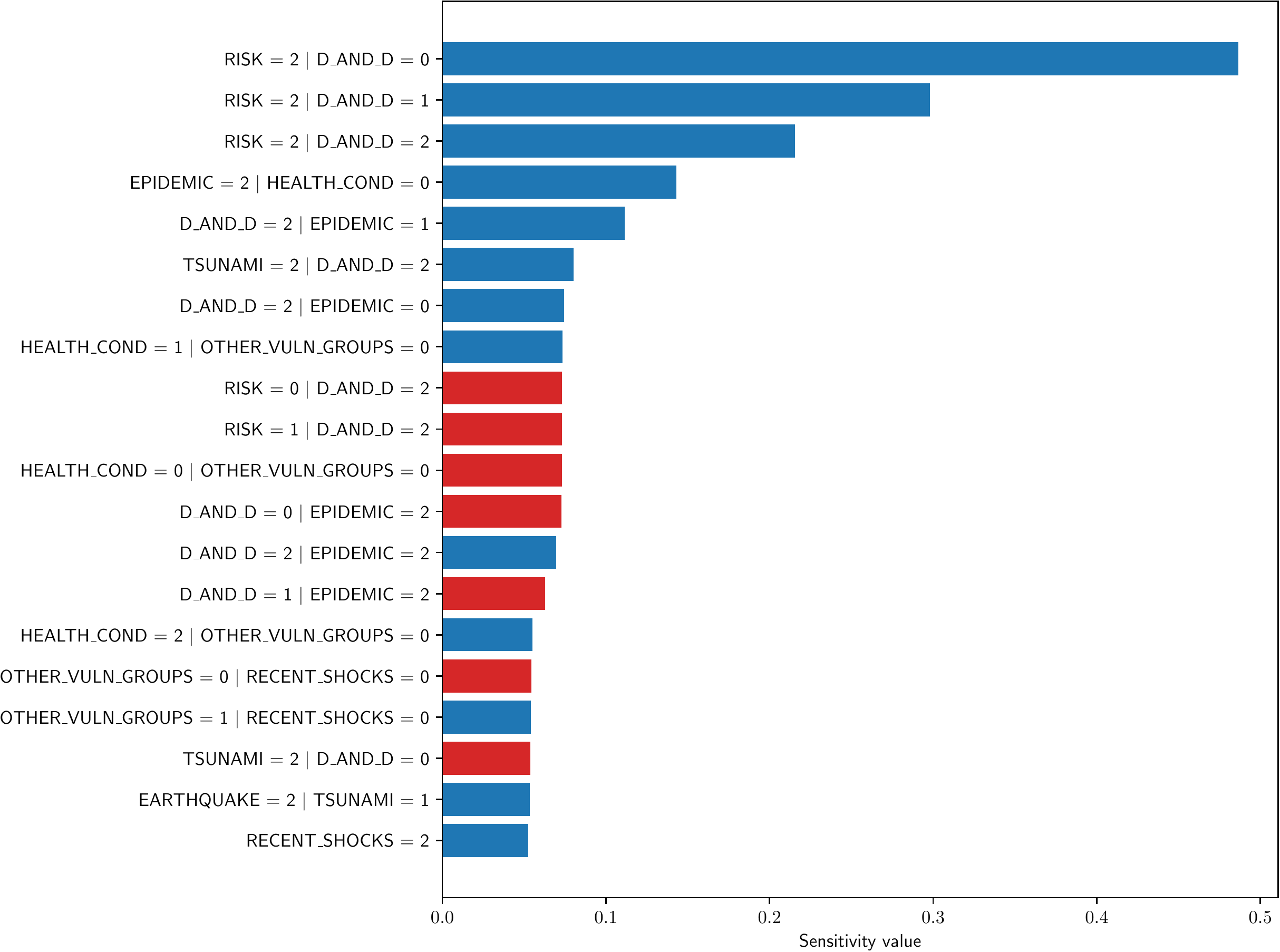}
		\vspace{-0.5cm}  
	\end{center}
	\caption{Top 20 most influential parameters for the humanitarian crisis network, color-coded by the sign of $f'(\theta_i)$. The probability of interest is $P(\mbox{RISK} = \mbox{high} \;|\; \mbox{EARTHQUAKE} = \mbox{high})$. Total computation time: 0.038s.}
	\label{fig:many_parameters_plot}
\end{figure}

As an additional illustration, Figure \ref{fig:many_parameters_plot} reports the sensitivity value of the parameters for the output conditional probability of an overall high risk given a high earthquake risk. Blue is associated with positive sensitivity values, and red with negative ones. Out of 183 network parameters, 30 have a sensitivity value of zero, meaning that they do not affect the probability of interest. It can be noticed that the most influential parameters have a positive relationship with the output probability, and almost all are associated with the development and deprivation index. 

\begin{figure}
	\begin{center}
		\includegraphics[scale=0.45]{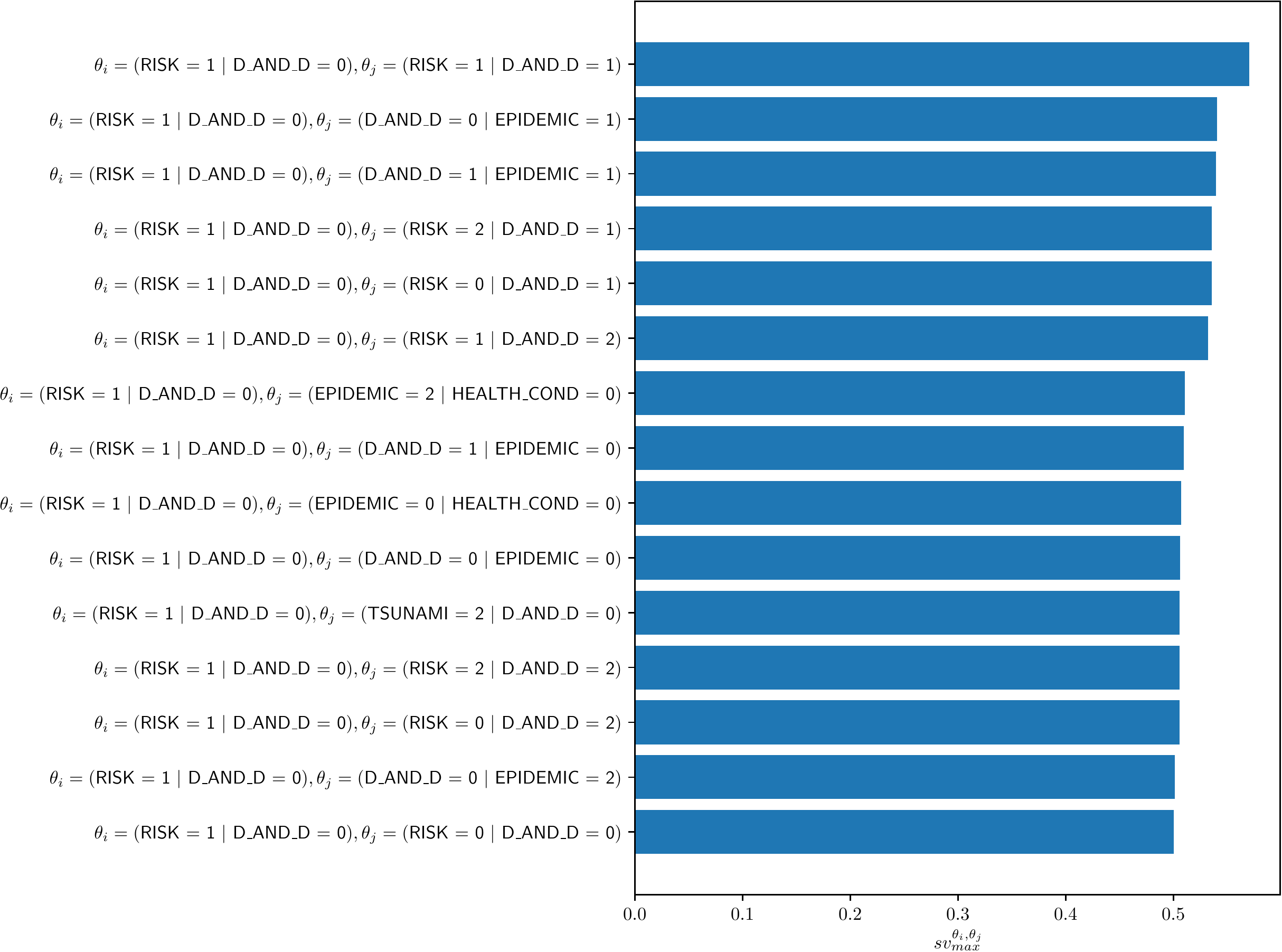}
		\vspace{-0.5cm}
	\end{center}
	\caption{Top 15 most influential pairs of parameters for the humanitarian crisis network according to $sv_{\mbox{max}}$. The probability of interest is $P(\mbox{RISK} = \mbox{high} \;|\; \mbox{EARTHQUAKE} = \mbox{high})$. Total computation time: 0.033s. \label{fig:max}}
\end{figure}

We further investigate in a 2-way sensitivity analysis the effect of parameters' variations over the same probability $P(\mbox{RISK} = \mbox{high} \;|\; \mbox{EARTHQUAKE} = \mbox{high})$. The 15 largest maximum 2-way sensitivity values are reported in Figure \ref{fig:max}. Since these are vector norms, they are always positive irrespective of the relationship between the parameters and the probability of interest. Thus, the coloring should not be interpreted as in Figure \ref{fig:many_parameters_plot}. Again all parameters associated with the development and deprivation index are the ones that have the most substantial effect on the probability of a country having a high overall risk. Thanks to the efficiency of YODO, these indices are almost instantaneously computed with a total computation time of just 0.047s. 

\subsection{The role of technology during COVID-19 isolation}

The second BN investigates the role of digital communication technology in facilitating the maintenance of meaningful social relationships and promoting the perception of social support during the COVID-19 lockdown. As reported by \citet{gabbiadini2020together}, the data was collected through an online questionnaire in March 2020 in Italy, about two weeks from the beginning of the lockdown that the Italian Government adopted for the urgent containment and management of the COVID-19 epidemiological emergency. The data can be downloaded from \citet{gab} and includes demographic information about 464 individuals, their use of digital communication technologies, and various psychological measures characterizing their emotional status. Each variable is discretized into either two, three, or four levels using either the equal frequency method or some ad-hoc thresholds to optimize the meaning of the levels. Details are reported in Table \ref{table:var2}.

\begin{table} \small
	\centering
 \scalebox{0.65}{
	\begin{tabular}{lrrr}
\toprule
 Variable & Meaning & Group &  Levels   \\
\midrule
\textbf{\cellcolor{gray!2}{AGE}}   &  \cellcolor{gray!2}{age of respondent}& \cellcolor{gray!2}{demographic}   &     \cellcolor{gray!2}{$<25$/$\geq 25$} (0/1)  \\
\textbf{\cellcolor{gray!2}{GENDER}}  & \cellcolor{gray!2}{gender of respondent} &   \cellcolor{gray!2}{demographic}   &     \cellcolor{gray!2}{male/female}(0/1) \\
\textbf{\cellcolor{gray!2}{REGION}}  & \cellcolor{gray!2}{region of residence}& \cellcolor{gray!2}{demographic}   &     \cellcolor{gray!2}{Lombardy/other}(0/1) \\
\textbf{\cellcolor{gray!2}{OUTSIDE}}  & \cellcolor{gray!2}{times outside per week}  & \cellcolor{gray!2}{demographic}   &     \cellcolor{gray!2}{0/1/$\geq$2}(0/1/2) \\
\textbf{\cellcolor{gray!2}{SQUARE\_METERS}}  & \cellcolor{gray!2}{home square meters} & \cellcolor{gray!2}{demographic}   &     \cellcolor{gray!2}{$<$80/$\geq$80}(0/1) \\
\textbf{\cellcolor{gray!2}{FAMILY\_SIZE}}  &   \cellcolor{gray!2}{number of individuals at home} &\cellcolor{gray!2}{demographic}   &     \cellcolor{gray!2}{1/2/$\geq3$}(0/1/2) \\
\textbf{\cellcolor{gray!2}{DAYS\_ISOLATION}}  &  \cellcolor{gray!2}{days since lockdown}  &\cellcolor{gray!2}{demographic}   &     \cellcolor{gray!2}{0-10/11-20/$>$20}(0/1/2) \\
\textbf{\cellcolor{gray!2}{OCCUPATION}}  & \cellcolor{gray!2}{occupation}  & \cellcolor{gray!2}{demographic}   &     \cellcolor{gray!2}{\specialcell{Other/Smartworking/Student/\\Office Work(0/1/2/3)}} \\
\textbf{\cellcolor{gray!5}{TECH\_FUN\_PQ}}  &   \cellcolor{gray!5}{\specialcell{use of communication technology \\for fun pre-quarantine}}
 &\cellcolor{gray!5}{technology}   & \cellcolor{gray!5}{low/medium/high}(0/1/2) \\
\textbf{\cellcolor{gray!5}{TECH\_FUN\_Q}}  &    \cellcolor{gray!5}{\specialcell{use of communication technology \\for fun  in quarantine}} & \cellcolor{gray!5}{technology}   &     \cellcolor{gray!5}{low/medium/high}(0/1/2) \\
\textbf{\cellcolor{gray!5}{TECH\_WORK\_PQ}}  &   \cellcolor{gray!5}{\specialcell{use of communication technology \\for work pre-quarantine}}
 &\cellcolor{gray!5}{technology}   & \cellcolor{gray!5}{low/high}(0/1) \\
\textbf{\cellcolor{gray!5}{TECH\_WORK\_Q}}  &    \cellcolor{gray!5}{\specialcell{use of communication technology \\for work  in quarantine}} & \cellcolor{gray!5}{technology}   &     \cellcolor{gray!5}{low/high}(0/1) \\
\textbf{\cellcolor{gray!8}{ANXIETY}}  & \cellcolor{gray!8}{\specialcell{level of anxiety}} &   \cellcolor{gray!8}{psychology}   &     \cellcolor{gray!8}{low/medium/high}(0/1/2) \\
\textbf{\cellcolor{gray!8}{ANG\_IRR}}  & \cellcolor{gray!8}{\specialcell{perceived level of anger/irritability}} & \cellcolor{gray!8}{psychology}   &     \cellcolor{gray!8}{low/medium/high}(0/1/2) \\
\textbf{\cellcolor{gray!8}{BELONGINGNESS}}  &  \cellcolor{gray!8}{\specialcell{how often the word we is used}} &\cellcolor{gray!8}{psychology}   &     \cellcolor{gray!8}{low/medium/high}(0/1/2) \\
\textbf{\cellcolor{gray!8}{BOREDOM}}& \cellcolor{gray!8}{\specialcell{level of boredom}}  &   \cellcolor{gray!8}{psychology}   &     \cellcolor{gray!8}{low/medium/high}(0/1/2) \\
\textbf{\cellcolor{gray!8}{LONELINESS}}  & \cellcolor{gray!8}{\specialcell{perceived loneliness}} & \cellcolor{gray!8}{psychology}   &     \cellcolor{gray!8}{low/medium/high}(0/1/2) \\
\textbf{\cellcolor{gray!8}{SOCIAL}}  &  \cellcolor{gray!8}{\specialcell{perceived social support}} &  \cellcolor{gray!8}{psychology}   &     \cellcolor{gray!8}{low/medium/high}(0/1/2) \\
\bottomrule
\end{tabular}}
	\caption{Variables considered for the COVID-19 network from \citet{gabbiadini2020together}.
}
	\label{table:var2}
\end{table}

\begin{figure}
    \centering
    \includegraphics[scale=0.5]{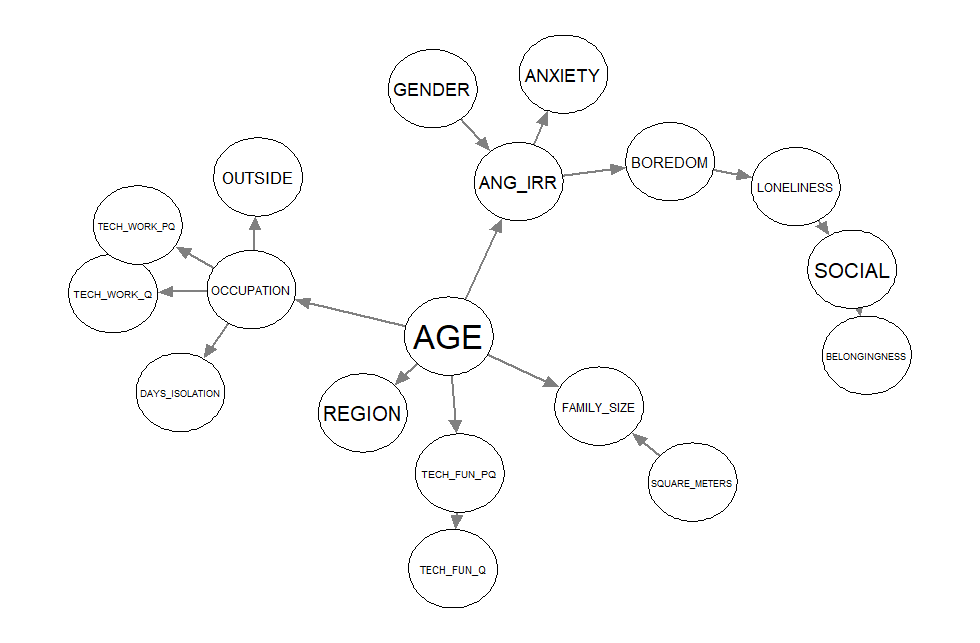}
    \caption{BN learned over the COVID-19 dataset from \citet{gabbiadini2020together}.
    \label{fig:bn2}}
\end{figure}

A BN is learned for this dataset using 1000 bootstrap repetitions of a tabu search algorithm and keeping the edges that have appeared more than 50\% of the times. Furthermore, edges from the psychological measures to the technological and demographic variables were forbidden. Similarly, no edges from the technological to the demographic variables were allowed. These choices were motivated by learning a network whose connections could have a more natural, causal interpretation. Figure \ref{fig:bn2} reports the learned BN. The two variables connected with psychological measures are age and gender. In particular, given the age of an individual, all other demographic characteristics (except gender) are irrelevant to predict his psychological status. The network, therefore, seems to suggest that age was the main driver for the psychological status of individuals in lockdown.

In this second application, we showcase the computation of the admissible regions using the YODO algorithm. Given a low level of the loneliness index, individuals were most likely spending much time interacting remotely for work during the quarantine (TECH\_WORK\_Q). Table \ref{tab:many_parameter_study1} reports the limits of the admissible regions and other measures ordered from the narrowest interval. The admissible region does not have width one in six cases, all coming from the pmf of TECH\_WORK\_Q or OCCUPATION. This suggests that the data strongly supports the hypothesis that individuals who did not feel lonely had many online work connections during the lockdown.

\begin{table} \tiny
	\centering
\scalebox{0.85}{
\begin{tabular}{llllll}
\toprule
{} &   Value &     Sens. value &            Proximity & AR 
(lower) & AR (upper) \\
Parameter                                          &         &  
&                      &            &            \\
\midrule
TECH\_WORK\_Q = 1 $\vert$ TECH\_WORK\_PQ = 1, OCCUPATION = 1 &  
$0.96$ &                $0.21$ &  $4.54 \cdot 10^{6}$ &     
$0.56$ &      $1.0$ \\
OCCUPATION = 0 $\vert$ AGE = 1                           &  
$0.37$ &                $0.28$ &  $9.91 \cdot 10^{6}$ &        
$0$ &     $0.67$ \\
OCCUPATION = 1 $\vert$ AGE = 1                           &   
$0.4$ &                $0.28$ &  $9.78 \cdot 10^{6}$ &      
$0.1$ &      $1.0$ \\
TECH\_WORK\_Q = 1 $\vert$ TECH\_WORK\_PQ = 0, OCCUPATION = 2 &  
$0.41$ &                $0.27$ &  $2.02 \cdot 10^{7}$ &    
$0.093$ &      $1.0$ \\
TECH\_WORK\_Q = 0 $\vert$ TECH\_WORK\_PQ = 0, OCCUPATION = 2 &  
$0.59$ &                $0.16$ &                  N/A &    
$0.068$ &      $1.0$ \\
TECH\_WORK\_Q = 1 $\vert$ TECH\_WORK\_PQ = 1, OCCUPATION = 2 &  
$0.75$ &                $0.12$ &  $5.49 \cdot 10^{7}$ &    
$0.041$ &      $1.0$ \\
AGE = 0                                            &  $0.45$ &  
$0.11$ &               $4.36$ &        $0$ &      $1.0$ \\
TECH\_FUN\_PQ = 1 $\vert$ AGE = 1                          &  
$0.26$ &   $4.0 \cdot 10^{-9}$ &   $2.1 \cdot 10^{7}$ &        
$0$ &      $1.0$ \\
TECH\_FUN\_PQ = 2 $\vert$ AGE = 0                          &  
$0.36$ &  $2.39 \cdot 10^{-8}$ &                  N/A &        
$0$ &      $1.0$ \\
TECH\_FUN\_PQ = 2 $\vert$ AGE = 1                          &  
$0.51$ &  $8.01 \cdot 10^{-9}$ &  $1.05 \cdot 10^{7}$ &        
$0$ &      $1.0$ \\
TECH\_FUN\_Q = 0 $\vert$ TECH\_FUN\_PQ = 0                   &  
$0.47$ &  $2.18 \cdot 10^{-8}$ &  $4.19 \cdot 10^{7}$ &        
$0$ &      $1.0$ \\
TECH\_FUN\_Q = 0 $\vert$ TECH\_FUN\_PQ = 1                   &  
$0.27$ &  $1.39 \cdot 10^{-8}$ &  $4.19 \cdot 10^{7}$ &        
$0$ &      $1.0$ \\
TECH\_FUN\_Q = 0 $\vert$ TECH\_FUN\_PQ = 2                   &  
$0.12$ &                   $0$ &                  N/A &        
$0$ &      $1.0$ \\
TECH\_FUN\_Q = 1 $\vert$ TECH\_FUN\_PQ = 0                   &  
$0.34$ &  $7.92 \cdot 10^{-9}$ &   $2.1 \cdot 10^{7}$ &        
$0$ &      $1.0$ \\
TECH\_FUN\_Q = 1 $\vert$ TECH\_FUN\_PQ = 1                   &  
$0.41$ &  $1.39 \cdot 10^{-8}$ &  $4.19 \cdot 10^{7}$ &        
$0$ &      $1.0$ \\
\bottomrule
\end{tabular}

}
	\caption{COVID network for the probability of interest $P(\mbox{TECH\_WORK\_Q} = \mbox{high} \;|\; \mbox{LONELINESS} = \mbox{low})$: sensitivity metrics for the 15 parameters with the smallest admissible region. Total computation time: 0.071s.}
	\label{tab:many_parameter_study1}
\end{table}

As a second illustration, we consider an individual's age, given that he felt very lonely during the lockdown. The BN suggests the most likely value was of individuals older than 24 years old. Table \ref{tab:many_parameter_study2} shows the admissible regions for the network parameters and shows that the network is way less robust for this hypothesis. Admissible regions are much narrower, having a width equal to 0.04 for two parameters. It can also be noticed that parameters with narrow admissible regions come from many different PMFs. Therefore, minor variations in the network parameters would make individuals with less than 25 years more likely to have high levels of loneliness. 

\begin{table} \tiny
	\centering
\scalebox{1}{
\begin{tabular}{llllll}
\toprule
{} &   Value & Sens. value & Proximity & AR (lower) & AR (upper)
\\
Parameter                           &         &                 
&           &            &            \\
\midrule
DAYS\_ISOLATION = 0 $\vert$ OCCUPATION = 3 &   $1.0$ &          
$0.028$ &   $11.13$ &     $0.96$ &      $1.0$ \\
OUTSIDE = 2 $\vert$ OCCUPATION = 3        &   $1.0$ &           
$0.028$ &   $11.13$ &     $0.96$ &      $1.0$ \\
ANG\_IRR = 2 $\vert$ AGE = 0, GENDER = 0   &  $0.27$ &          
$0.047$ &     $8.4$ &        $0$ &     $0.29$ \\
BOREDOM = 0 $\vert$ ANG\_IRR = 1           &  $0.24$ &          
$0.019$ &     $3.4$ &        $0$ &      $0.3$ \\
ANG\_IRR = 0 $\vert$ AGE = 1, GENDER = 1   &   $0.3$ &          
$0.13$ &    $2.43$ &        $0$ &     $0.31$ \\
LONELINESS = 1 $\vert$ BOREDOM = 0        &  $0.33$ &           
$0.014$ &     $7.4$ &        $0$ &     $0.41$ \\
LONELINESS = 0 $\vert$ BOREDOM = 1        &  $0.32$ &           
$0.011$ &    $4.28$ &        $0$ &     $0.42$ \\
AGE = 1                             &  $0.55$ &            
$1.01$ &   $0.023$ &     $0.54$ &      $1.0$ \\
AGE = 0                             &  $0.45$ &            
$0.59$ &     $1.3$ &        $0$ &     $0.46$ \\
GENDER = 1                          &  $0.75$ &          
$0.0048$ &     $4.6$ &     $0.52$ &      $1.0$ \\
GENDER = 0                          &  $0.25$ &          
$0.0048$ &     $4.6$ &        $0$ &     $0.48$ \\
ANG\_IRR = 1 $\vert$ AGE = 1, GENDER = 1   &  $0.42$ &          
$0.018$ &   $23.86$ &        $0$ &     $0.49$ \\
ANG\_IRR = 2 $\vert$ AGE = 0, GENDER = 1   &  $0.49$ &          
$0.13$ &    $2.42$ &        $0$ &      $0.5$ \\
LONELINESS = 1 $\vert$ BOREDOM = 1        &  $0.47$ &           
$0.011$ &    $4.28$ &        $0$ &     $0.57$ \\
BOREDOM = 1 $\vert$ ANG\_IRR = 1           &   $0.5$ &          
$0.016$ &    $4.12$ &        $0$ &     $0.58$ \\
\bottomrule
\end{tabular}

}
	\caption{COVID network for the probability of interest $P(\mbox{AGE} = \geq 25 \;|\; \mbox{LONELINESS} = \mbox{high})$: sensitivity metrics for the 15 parameters with the smallest admissible region. Total computation time: 0.11s. }
	\label{tab:many_parameter_study2}
\end{table}


\section{Discussion}
\label{sec:discussion}

We demonstrated the use of automatic differentiation in BNs and, more specifically, in studying how sensitive they are to parameter variations. The novel algorithms are freely available in Python and are planned to be included in the next release of the \texttt{bnmonitor} R package \citep{leonelli2021sensitivity}. Their efficiency was demonstrated through a simulation study. Two critical applications in humanitarian crises and studying the psychological effects of isolation during the COVID-19 pandemic illustrate their use in practice.

Although YODO is specifically designed to compute the coefficients of the one-way sensitivity function in Equation \ref{eq:sens}, we demonstrated in this paper how it could be used to answer a variety of sensitivity queries, for instance, admissible regions and the identification of the parameter sensitivity set. Importantly, YODO also provides the basis for multi-way sensitivity analyses, and we demonstrated their feasibility in practice. 

\subsection*{Future Work}

The YODO algorithm introduced here is designed explicitly for BN models, but it could also be adapted to work with other graphical models. The study of context-specific independence has been shown to increase the efficiency of various inferential tasks often, and thus we may expect that it could also speed up YODO. Therefore, we plan to adapt it to work over graphical models embedding non-symmetric types of independence, as, for instance, staged trees \citep{Carli2020,smith2008conditional}, whose sensitivity functions have also been studied \citep{leonelli2019sensitivity}. Another avenue of research is the adaptation of YODO to work for sum-product networks \citep{poon2011sum,sanchez2021sum}, a different representation of a factorization of a joint probability distribution, which has become increasingly popular in the past few years.

Although YODO makes various types of multi-way sensitivity analysis feasible, they are still a local approach to investigate the combined effect of parameters' variations on probabilities of interest. Recently, it has been shown that the computation of Sobol indices, a global sensitivity index, is feasible in sensitivity to evidence analyses \citep{ballester2022computing}.  We are currently investigating algorithms to globally assess the effect of the various parameters of a BN and consequently compute their associated Sobol indices. 



\bibliography{references}





\end{document}